\documentclass[aps,prb,twocolumn,floats,graphicx,epsfig,euscript,amsmath,amssymb]{revtex4}
\usepackage[cp1251]{inputenc}
\usepackage{mathrsfs}
\usepackage[dvips]{graphicx}
\usepackage{graphics}
\usepackage{color}
\usepackage{epsfig}
\DeclareMathAlphabet\EuScript{U}{eus}{m}{n} \SetMathAlphabet\EuScript{bold}{U}{eus}{b}{n}

\def\lapprox{\,\raise0.4ex\hbox{$<$}\kern-0.8em\lower0.7ex\hbox{$\sim$}\,}
\def\gapprox{\,\raise0.4ex\hbox{$>$}\kern-0.8em\lower0.7ex\hbox{$\sim$}\,}
\def\be{\begin{equation}}
\def\ee{\end{equation}}
\def\ba{\begin{array}}
\def\ea{\end{array}}

\def\bc{\begin{center}}
\def\ec{\end{center}}

\begin{document}
\bibliographystyle{prsty}
\title
{Spin-flip excitations and Stoner ferromagnetism in a strongly correlated quantum Hall system }

\author{S. Dickmann$^1$ and B.$\,$D. Kaysin$^{1,2}$}

\address{$^1$Institute of Solid State Physics, Russian
Academy of Sciences, Chernogolovka, 142432 Russia\\
$^2$Moscow Institute of Physics and Technology, Dolgoprudny 141700, Russia}


\begin{abstract}
Spin-flip excitations in a quantum Hall electron system at fixed filling factor $\nu=2$ are modelled and studied under conditions of a strong Coulomb interaction when the `Landau level mixing' is a dominant factor determining the excitation energy. The `one-exciton' approach used for the purely electronic excitations in question allows us to describe the Stoner transition from the unpolarized/paramgnet state to the polarized/ferromagnet one. The theoretical results are compared with the available experimental data.
\end{abstract}
\maketitle

\section{Introduction}
\vspace{-2mm}

Most quantum-Hall (QH) systems created in modern materials, for instance, graphene and
MgZnO/ZnO structures, are characterized by large Wigner-Seitz parameter $r_s$ -- ratio of characteristic Coulomb energy $e^2/\kappa l_B(\sim 10\!-\!20\,$meV) to relevant single electron energy $\hbar\omega_c$. (Here $\kappa$, $l_B$ and $\omega_c$ are the static dielectric constant, magnetic length and cyclotron frequency, respectively.) The strong Coulomb correlation should inevitably result in essential `mixing' of different Landau levels. Such a feature, however, does not smooth out, but rather strengthens the quantum Hall properties of the system. Indeed, experimentally, the two-dimensional (2D) electron plasma in a perpendicular magnetic field in MgZnO/ZnO heterostructures represents a typical QH system where the characteristic properties are clearly manifested: for example, there is a sharp dependence of the magneto-transport on the value of the $\nu\sim 1$ filling factors.\cite{fa15} This means that at least in the ground state large $r_s$ (in MgZnO/ZnO we have typical range $7\!<r_s\!<10$) does not result in substantial smearing of electron density over a large number of Landau levels. Another interesting fact is that by now all the theoretical studies of QH systems have been based on the formal assumption of $r_s$ smallness (see, for instance, Refs, \onlinecite{la83}, and \onlinecite{ka84}), this approach being often fairly successful. Parameter $r_s$ under real experimental conditions is still of the order of one even in GaAs/AlGaAs quantum wells. However, the theory advances in microscopic description of QH systems (for example, the very accurate description given by R.B. Laughlin$\,$\cite{la83}  of some fractional QH states by using a combination of single-electron wave functions of only the Landau zero level) indicate that the Coulomb mixing of different Landau levels is often effectively very small even if $r_s\sim 1$. Kohn's theorem$\,$\cite{ko61}  also points to some hidden relationship between the single-particle Landau states in the magnetic field and the Coulomb interaction, preserving, in a sense, the hierarchy of Landau levels regardless of the magnitude of the interaction, i.e. at any $r_s$.

In the present work we study lowest-energy excitations in a strongly correlated system as applied to the case of 2D plasma in a MgZnO/ZnO heterostructure where, certainly, parameter $r_s$ can in no way be considered a small value. The QH system with filling factor $\nu=2$ is modelled with the help of an approach based on experimental data and some general assumptions. We find that breakdown of the spin-unpolarized/paramagnet phase and Stoner transition to the spin-polarized/ferromagnet state takes place due to short-wave (with wave-length ${}\!\sim\!2\pi l_B\!/1.4$) thermal single-spin-flip fluctuations, when their energy vanishes or in fact becomes lower than temperature $T$ ($\simeq 0.5\,$K). We do not consider backward transition from the ferromagnet to paramagnet phase, however, in the final part we discuss spin excitations in the $\nu\!=\!2$ ferromagnet and the probable mechanism of `reverse' Stoner transition.

Below in our calculations the energy is everywhere measured in $e^2\!/\kappa l_B$ Coulomb units, so the dimensionless cyclotron and Zeeman gaps are
$w_c=r_s^{-1}=\hbar\omega_c\!/(e^2\!/\kappa l_B)$ and $e_{\rm Z}=g\mu_BB\!/(e^2\!/\kappa l_B)$
respectively. The numerical values are\vspace{-1mm}
\begin{equation}\label{gaps}\vspace{-1mm}
w_c=0.057\sqrt{B}\quad {\rm and}\quad e_{\rm Z}=0.017\sqrt{B},
\end{equation}
if $B$ is measured in tesla.

\section{Formalism of the excitonic representation}
\vspace{-2mm}

We present formalism describing the electron QH system by using the so-called excitonic representation (for more details see Refs. \onlinecite{di05} and \onlinecite{di19}). The main idea of the excitonic representation is to abandon the basis of Fermi one-electron states and switch to the basis of so-called exciton states that diagonalize some essential part of the Coulomb interaction. The exciton states in a purely electronic QH system are generated by operators originally defined via Dirac electron operators: if $p$ is the undimensionalized  (in $1/l_B$ units) `intrinsic' quantum number of a continually degenerated Landau level and $a_p$ and $b_p$ are annihilation operators corresponding to binary indexes $a$ and $b$ [each designates both the Landau level number and the spin sublevel, $a=(n_a,\sigma_a)$], then the exiton creation operator is\vspace{-0.mm}
\begin{equation}\label{Q}
\displaystyle{{\cal Q}_{ab\,{\bf q}}^{\dag}={{\cal N}_{\phi}^{-1/2}}\sum_{p}\,
  e^{-iq_x\!p}\,
  b_{p+\frac{q_y}{2}}^{\dag}\,a_{p-\frac{q_y}{2}}},\vspace{-0.5mm}
\end{equation}
where ${\cal N}_{\phi}$ is the number of magnetic flux quanta in the system in question. \vspace{-1.mm}$\left(\mbox{Note also that}\, {\cal Q}_{ab\,{\bf q}}\!\equiv {\cal Q}_{ba-\!{\bf q}}^{\dag}\right)$. These ${\cal Q}$-operators have a very important property: when acting on the state of the QH system they add value $\hbar{\bf q}/l_B$ to total momentum of the system since there occurs commutator equality \vspace{-1.mm}$\left[{\hat P},{\cal Q}_{ab{\bf q}}^{\dag}\right]={\bf q}{\cal Q}_{ab{\bf q}}^{\dag}$ [where ${\hat P}$ describes the dimensionless (with $\hbar=l_B=1$) `momentum' operator$\,$\cite{di19}]. In particular, if $|0\rangle$ is the ground state, then the exciton state ${\cal Q}_{ab\,{\bf q}}^{\dag}|0\rangle$, \vspace{-.5mm} if not zero, is the eigenstate of momentum operator ${\hat P}$ with  eigen quantum number ${\bf q}$. Thus, exciton states, in contrast to single electron states, possess a natural quantum number, namely the 2D momentum whose existence is the consequence of the translational invariance of the QH system.

The general expression of the total Coulomb Hamiltonian of the 2D electron system can be presented in terms of the excitonic ${\cal Q}$-operators:\cite{di05,di19}\vspace{-1.mm}
\be\label{Q-presentation}\begin{array}{l}
\displaystyle{{\hat H}_{\rm Coul}\!=\!\!{}\!{}\!\!{}\!{}\!{}\!\sum_{{\bf q},a,b,c,d}\!\!\!\!\frac{{\cal F}\!(q)}{2q}
  \left(h_{n_an_b{\bf q}}\delta_{\sigma_a,\sigma_b}{\cal Q}_{a\!b{\bf q}}^{\dag}\right)}\,\quad\vspace{-1.mm}\\{}\qquad{}\qquad{}\qquad{}\quad
\times\!\displaystyle{\left({h_{n_cn_d-{\bf q}}\delta_{\sigma_c,\sigma_d}{\cal Q}}_{cd\,-\!{\bf q}}^{\dag}\right)}\vspace{1.mm}\\
  \qquad{}\qquad{}\qquad\displaystyle{-\sum_{{\bf q},a,b}
  \!\frac{{\cal F}_{ee}\!(q)}{2q}
  \left|h_{n_an_b}({\bf q})\right|^2{\cal B}^{\dag}_0}
\end{array}\vspace{-1.mm}
\ee
In this equation ${\cal B}_0^\dag$ is the ${\bf q}\!=\!0$ `intra-sublevel' operator:
$\displaystyle{{\cal B}_{{\bf q}}^{\dag}\equiv{N_{\phi}^{-1/2}}{\cal Q}_{bb{\bf q}}^\dag}\;\,$
$\left[{\cal B}_{{\bf q}}={\cal B}_{-{\bf q}}^{\dag}\right]$; ${\cal F}(q)$ is  the effective formfactor:\cite{an82}
\vspace{-1.mm}
\begin{equation}\label{formfactor}
{}\!\!{\cal F}(q)\!\displaystyle{=\!\int\!\!\!\int\!
dz_1dz_2e^{-q|z_1-z_2|/l_B}|\chi(z_1)\chi(z_2)|^2\!,}\vspace{-1.mm}
\end{equation}
where  $\chi(z)$ describes the electron size-quantized function in the quantum well; the $h$-functions, the factors at the ${\cal Q}$-operators in Eq. \eqref{Q-presentation}, are\vspace{-1.mm}
\be\label{h-functions}
  h_{kn{\bf q}}=\left(\frac{k!}{n!}\right)\!\!{\vphantom{\left(\frac{k!}{n!}\right)}}^{1/2}\!\!e^{-q^2/4}
  (q_-)^{n\!-\!k}L^{n\!-\!k}_k(q^2\!/2) \vspace{-1.mm}
\ee
[$q_{\pm}=\pm\frac{i}{\sqrt{2}}(q_x\pm iq_y)$, $L_n^k$ is the Laguerre polynomial and $\delta_{...,...}$ is Kronecker delta]. They satisfy identity $h_{kn{\bf q}}\equiv h^*_{nk\,-{\bf q}}$. The one-particle part of the Hamiltonian is presented by cyclotron and Zeeman terms which in the excitonic representation are\vspace{-1.mm}
\be\label{H^1}\begin{array}{l}
\!{}\!\displaystyle{{\hat H}^{(1)}\!=\!w_c{\cal N}_\phi\sum_a\!\left(n_a+\frac{1}{2}\right)\!{\cal A}_0}\\
  \displaystyle{\quad\qquad-\frac{1}{2}\,e_z{\cal N}_\phi\sum_a{\cal A}_0\left(\delta_{\sigma_a,\uparrow}-\delta_{\sigma_a,\downarrow}\right)}
  \end{array}\vspace{-1.mm}
\ee
(see the definition of operator ${\cal B}_{\bf q}$ above; ${\cal A}_{\bf q}$ means the $b\to a$ replacement), where we consider the positive spin is regarded as in the opposite direction to the magnetic field.

Certainly, the ${\cal Q}$-operators \eqref{Q} do not belong to Bose or Fermi types, they form a proper Lie algebra with commutation rules \vspace{-1.mm}
\begin{equation}\label{commutators}\begin{array}{l}
{}\!{}\!{}\!{}\!{}\!{}\!\left[{\cal Q}_{cd\,{\bf q}_1}^{\dag}\!,\!{\cal Q}_{a\!b\,{\bf
   q}_2}^{\dag}\right]\!\equiv\! {\cal N}_{\phi}^{-1/2}\!\!\left( e^{-i{\bf
   q}_1\!\times{\bf q}_2/2}\delta_{b,c}{\cal Q}_{a\!d\,{\bf q}_1\!+\!{\bf
   q}_2}^{\dag}\right.\vspace{2mm}\\\qquad\qquad{}\qquad\qquad\left.-e^{i{\bf
   q}_1\!\times{\bf q}_2/2}\delta_{a,d}
   {\cal Q}_{cb\,{\bf q}_1\!+\!{\bf
   q}_2}^{\dag}\right). \vspace{-1.mm}
\end{array}
\end{equation}
For a fixed pair of different indexes $(a,b)$ we have\vspace{-1.mm}
$$
\left[{\cal Q}_{a\!b{\bf q}_1}^\dag,{\cal Q}_{a\!b{\bf q}_2}^\dag\right]\!=\!\left[\vphantom{{\cal Q}_{a\!b{\bf q}_1}^\dag}{\cal Q}_{a\!b{\bf q}_1},{\cal Q}_{a\!b{\bf q}_2}\right]\equiv 0, \vspace{-1.mm}
$$
and\vspace{-1.mm}
\begin{equation}\label{commutator_ab}
\left[{\cal Q}_{a\!b\,{\bf q}\!{}_1},\!{\cal Q}_{a\!b{\bf q}_2}^\dag\right]\!\!
=\!e^{i{\bf q}\!{}_1\!\times{\bf q}_2\!/2}\!\!{\cal A}_{{\bf q}\!{}_1-{\bf q}_2}\!-
e^{i{\bf q}_2\!\times{\bf q}\!{}_1/2}{\cal B}_{{\bf q}\!{}_1\!-{\bf q}_2},\vspace{-1.mm}
\end{equation}
where $a\!\neq\!b$. Besides,\vspace{-0.mm}
\begin{equation}\label{commA}
\begin{array}{l}
\displaystyle{{}\!{}\!{}\!e^{i{\bf q}_1\!\times{\bf q}_2/2}[{\cal A}_{{\rm q}_1},
  {\cal Q}^\dag_{ab{\rm q_2}}]\!=\!
  -e^{-i{\bf q}_1\!\times{\bf q}_2/2}[{\cal B}_{{\rm q}_1},
  {\cal Q}^\dag_{ab{\rm q_2}}]}\vspace{1.mm}\\\qquad\qquad{}\qquad{}\qquad\qquad\displaystyle{=
  -{\cal N}_\phi^{-1}{\cal Q}^\dag_{ab\,{\rm q}_2\!-{\rm q}_1}}
\end{array}\vspace{-0.mm}
\end{equation}

Let $|0\rangle$ be the ground state of an integer QH system where the sublevel $a$ is completely occupied and the sublevel $b$ is completely empty. Then we have ${\cal A}_{\bf q}|0\rangle=\delta_{{\bf q},0}$ and ${\cal B}_{\bf q}|0\rangle\equiv 0$, and commutator \eqref{commutator_ab}, if averaged over the ground state, represents a common permutation identity for Bose states. As a result, exciton states ${\cal Q}_{ab}^\dag|0\rangle$ {\em obey Bose-Einstein statistics} despite being collective excitations in a QH fermionic (purely electronic) system. If all ${\cal Q}$-excitations were usual Bose particles, then the first two-operator term of the Coulomb Hamiltonian \eqref{Q-presentation} would represent a combination of different components of their density operator, where, after appropriate diagonalization, an inter-particle coupling might be effectively excluded. Generally, this does not occur: in particular, the action of the Coulomb operator on a single-exciton state provides, due to the exact commutation rules \eqref{commutators}, a quantum fluctuation to a double-exciton state, although with total momentum preserved.

\section{The one-exciton model used to describe excitations from the unpolarised ground state}
\vspace{-2mm}

We formulate the properties of the model used for calculating the spectrum of spin-flip excitations from the ground state. Recall that we study an $\nu=2$ spin-unpolarized QH system. The set of various single-electron states of Landau levels is quite complete. It is obvious that the $N_e$-electron wave function can always be represented as a combination of $N_e$-fold products of one-electron functions corresponding to states of degenerated Landau levels. The simplest way to `arrange' $N_e=2{\cal N}_\phi$ electrons with total spin $S=0$ is {\em to model the ground state} $|0\rangle$ {\em by a fully occupied zeroth Landau level}. Thus, even taking into account that the electronic functions of the lowest Landau-level should be renormalized due to the strong Coulomb interaction, we assume that the structure of the ground state of this Fermi system remains basically the same as in the absence of an interaction (see the above discussion concerning  features of quantum Hall systems).

Besides, to support the chosen approach one can roughly analyze the situation where the ground state is modelled by Landau-level partial fillings $\nu_a$ satisfying the total condition $\sum_a\nu_a=2$, and the occupied $p$-states on every degenerated Landau level are uniformly distributed. Then, when calculating ground-state energy within the framework of the Hartree-Fock approach,  we come to the following result: depending on magnetic field the energy minimum is always reached at integer partial filling factors, namely, either at $\nu_0=\nu_{\overline 0}=1$ (the unpolarized state) or at $\nu_{0}=\nu_{1}=1$ (the ferromagnet state). [Here and below we use new notations for the spin-sublevel indexes: $n=(n,\uparrow)$ and ${\overline n}=(n,\downarrow)$.]

Excitations from the unpolarized state corresponding to the $S=1$ spin, represent a triplet with $S_z=-1,0,1$ where the spin-components at any fixed ${\bf q}$ are energetically equidistant and separated by the Zeeman gap. So, it is sufficient to study only one lowest component with $S_z=1$. It is remarkable that the previous calculations performed in terms of the $r_s$-expansion show that the minimum energy of this excitation is in the vicinity of finite $q=q_0\simeq 1$ resulting in a gap that is narrower than the single-electron value $\hbar\omega_c-g\mu_BB$.\cite{ka84} Besides, the gap tends to decrease with the growth of $r_s$ (with magnetic field weakening).

The energy should be counted from the ground state level $\langle 0|{\cal H}|0\rangle$ where
\begin{equation}\label{Hamiltonian_tot}
  {\cal H}={\hat H}^{(1)}+{\hat H}_{\rm Coul}
\end{equation}
is the total Hamiltonian. An essential feature of our model consists in a limitation of the basic-set for the $S=S_z=1$ excitations: assuming that the excitation with the lowest energy should be arranged in the simplest way, we will consider it {\em only within the framework of the one-exciton approach}.
That is, as basis states, only single-exciton states ${\cal Q}_{{ab}{\bf q}}^\dag|0\rangle$ are used, and for the spin-flip excitations: $a=(0,\downarrow)\!\equiv {\overline 0}$ and $b=(n,\uparrow)\!\equiv n$ with $n>0$.
Generally, it is a reduced basic-set. We ignore, for example, double-exciton states where the spin-flip mode occurs along with a magnetoplasma one: namely, states  of type ${\cal Q}_{{0m}{\bf q}_1}^\dag{\cal Q}_{{\overline 0}n\,{\bf q}_2}^\dag|0\rangle$ or ${\cal Q}_{{\overline 0}{\overline m}\,{\bf q}_1}^\dag\!{\cal Q}_{{\overline 0}n\,{\bf q}_2}^\dag|0\rangle$, where $m=1,2,3,...$ and $n=1,2,3,...$.

Another interpretation of our model can be formulated as follows: in the Hamiltonian ${\cal H}$ we keep only the excitonically diagonalizable part ${\cal H}_{\rm ED}$ which, acting onto a basis state ${\cal Q}_{{\overline 0}n{\bf q}}^\dag|0\rangle$, results in a combination of basis one-exciton states: $\sim \sum_{m}C(m,{\bf q}){\cal Q}_{{\overline 0} m{\bf q}}^\dag|0\rangle$. (The same spin and momentum quantum numbers are preserved due to the properties of the total Hamiltonian.) The operator ${\hat H}^{(1)}$ and one-exciton terms ($\sim\!\!\!\sum_b\!...{\cal B}_0$) in ${\hat H}_{\rm Coul}$ are definitely included in ${\cal H}_{\rm ED}$. The other `non-diagonalizable' members of the Hamiltonian ${\cal H}_{\rm non-ED}={\cal H}-{\cal H}_{\rm ED}$,  when acting onto the basis state, result, for instance, in two-exciton states (or even three-exciton ones) with additional magnetoplasma modes. Projection of these states onto any single spin-flip exciton is equal to zero due to vanishing three-operator expectation, $\langle 0|{\cal Q}_{{\overline 0} k{\bf q}}{\cal Q}_{0{m}{\bf q}-{\bf q}'}^\dag{\cal Q}_{{\overline 0} n{\bf q}'}^\dag|0\rangle\!\equiv 0$, which occurs at any nonzero numbers $n,m$ and $k$. At the same time, basically, the energy of two-exciton state $|m,n,{\bf q},{\bf q}'\rangle={\cal Q}_{0{m}{\bf q}-{\bf q}'}^\dag{\cal Q}_{{\overline 0}n{\bf q}'}^\dag|0\rangle$, if inter-excitonic coupling is neglected, is determined by the expectation $\langle {\bf q}',{\bf q},n,m|[{\cal H}, {\cal Q}_{0{m}{\bf q}-{\bf q}'}^\dag{\cal Q}_{{\overline 0} n{\bf q}'}^\dag]|0\rangle$ (energy is considered to be counted from the ground state level), and hence, it is the sum of energies of the spin-flip and magnetoplasma modes. The previous calculations$\,$\cite{ka84}, show that at $q\sim 1$ the energy of the magnetoplasma mode is significantly higher than that of the spin-flip mode. The reason is related to the exchange energy arising from the terms of the Hamiltonian \eqref{Q-presentation}, where the electron-`hole' pair constituting the magnetoplasma exciton annihilates at one point in the $K$-space and is simultaneously created at another point. This exchange contribution is absent in the case of a spin-flip exciton.\cite{ka84} Thus, there is
an argument in favor of the chosen model: the two-exciton states at $q\sim 1$ are energetically distant from the spin-flip one-exiton states.\cite{foot1}

\begin{figure}[h]
	\vspace{-4.mm}
	\hspace{-0.mm}
	\begin{center}
		\includegraphics*[angle=0,width=.47\textwidth]{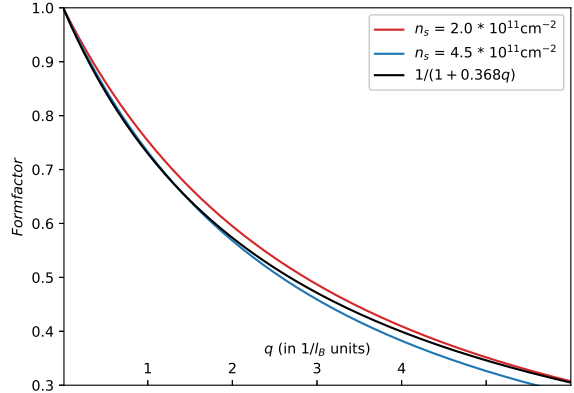}
	\end{center}
	\vspace{-2.mm}
	\caption{Formfactor ${\cal F}(q)$ for different electron densities (see text).}
\end{figure}

Before presenting the results of the calculations, one further comment should be made on the formfactor used. The latter is found within an approach repeatedly used in previous works.\cite{pi,ma13} The electron motion in the $z$-direction is assumed to be independent of the magnetic field but governed, except the external potential, by the $e$-$e$ Coulomb interaction. The single-electron function $\chi(z)$ in the expression \eqref{formfactor} is calculated self-consistently and becomes dependent on the electron density. When the filling factor is fixed, the formfactor turns out to be dependent on the magnetic field not only artificially (due to the presentation of the wave vector ${\bf q}$ in $l_B^{-1}$ units) but parametrically through the dependence of $\chi$ on electron density. It is interesting that both dependencies compensate each other well. The formfactors ${\cal F}(q)$ calculated for quantum wells$\,$\cite{ko14} for different electron densities, $n_s=4.9\!\times\!10^{10}(B/{\rm tesla}){\rm cm}^{-2}$, are shown graphically in Fig. 1 as functions of wave vector $q$ measured in units of appropriate values of $l_B^{-1}=3.9\!\times\! 10^{5}\sqrt{B/{\rm tesla}}\,\,{\rm cm}^{-1}$. These curves are fairly close to each other, and as a fitting function, we use the simple formula ${\cal F}(q)=(1+0.368q)^{-1}$ (see the black line in Fig. 1). So, the  Coulomb part of the Hamiltonian, if measured in Coulomb units, becomes independent of $B$. The dependence, however, is present in operator ${\hat H}^{(1)}$ [see Eqs. \eqref{H^1} and \eqref{gaps}].

\section{Results}
\vspace{-2mm}

Now we find the matrix elements of the Hamiltonian \eqref{Hamiltonian_tot}:
${\cal M}_{nm}(q)=
\langle {\cal Q}_{0\overline{m}}|[{\cal H},{\cal Q}_{0\overline{n}}^\dag] \rangle $
($m,n=1,2,3,...$; the angular brackets $\langle...\rangle$ mean averaging over the ground state $|0\rangle$; here and elsewhere below we, though assuming, but omit the `${\bf q}$' subscript at ${\cal Q}$  in the angle brackets). This calculation is performed with the help of formulae \eqref{Q-presentation}-\eqref{commA}, and represents a routine algebraic manipulation. As a result, in terms of $h$-functions we obtain diagonal matrix elements\vspace{-1.mm}
\begin{equation}\label{elements}
\ba{l}
\!\!\!\displaystyle{{\cal M}_{nn}(q)=\Delta_n+\sum_{{\bf q}_1}\frac{{\cal F}(q_1)}{q_1{\cal N}_\phi}\left[{h_{00}(q_1)}^2-|h_{0n}({\bf q}_1)|^2\right.}\vspace{-0.mm}\\
{}\qquad{}\qquad{}\qquad\left.-e^{i{\bf q}\!\times{\bf q}_1}\,{h_{00}(q_1)}{h_{nn}(q_1)}\vphantom{{{\cal M}_{nn}(q)=\sum_{{\bf q}_1}\frac{{\cal F}(q_1)}{q_1{\cal N}_\phi}\left[{h_{00}(q_1)}^2-|h_{0n}({\bf q}_1)|^2\right.}}\right],\vspace{-1.mm}
\ea
\end{equation}
where $\Delta_n\!=\!nw_c-e_{\rm Z}.$ Calculating the non-diagonal elements we note that the sum
${\cal N}_\phi^{-1}\sum_{m'\!,{\bf q}}\left[{\cal F}(q)/q\right]h_{nm'}({\bf q})h^*_{mm'}({\bf q})$~vanishes~if $n\!\neq\!m$, and then obtain: \vspace{-1.mm}
\begin{equation}\label{non_diag}
\displaystyle{{\cal M}_{nm}(q)\!=\!-\!\!\sum_{{\bf q}_1}\!\frac{{\cal F}(q_1)}{q_1{\cal N}_\phi}\,{h_{00}(q_1)h_{nm}({\bf q}_1)\,e{}\!{}\!{}^{-i{\bf q}_1\!\times{\bf q}}}}\vspace{-1.mm}
\end{equation}
(here $m\!\neq\!n$). Using Eq. \eqref{h-functions} and performing summations, $\sum_{\bf q}...={\cal N}_\phi\int...qdqd\varphi/2\pi$, we get\vspace{-1.mm}
\begin{equation}\label{elements2}\ba{l}
{}\!{}\!\displaystyle{{\cal M}_{nn}\!\!=\!\Delta_n\!+\!\!\int_0^\infty\!\!\!\!\!dq_1{\cal F}(q_1)e^{-q_1^2/2}\!\left[1\!-\!\frac{1}{n!}{\left(\frac{q_1^2}{2}\right)^n}\right.}\vspace{1.mm}\\
\qquad{}\qquad{}\qquad{}\qquad\displaystyle{-L_n({q_1^2}/{2})J_0(qq_1)\mbox{\huge{]}},}
\ea\vspace{-1.mm}
\end{equation}\vspace{-1.mm}
and\vspace{-1.mm}
\begin{equation}\label{non_diag2}\ba{l}
\displaystyle{{\cal M}_{n\neq m}=-i^{m-n}\,\sqrt{\frac{n!}{m!}}\int_0^\infty\!\!\!\!dq_1{\cal F}(q_1)e^{-q_1^2/2}}\vspace{1mm}\\
\displaystyle{\qquad\times\,\left(\frac{q_1}{\sqrt{2}}\right)^{m-n}\!\!L_n^{m-n}({q_1^2}/{2})
J_{m-n}(qq_1),}
\ea
\end{equation}
where $J_{n}(x)$ is the Bessel function. Within the framework of our approach we come to
the secular equation\vspace{-1.mm}
\begin{equation}\label{secular}
\left|{\cal M}_{nm}(q)-E(q)\delta_{n,m}\right|, \quad n,\, m = 1,2,3,...\vspace{-1.mm}
\end{equation}
Taking into account the restricted application of our model, we will limit ourselves only to considering the lowest-energy root. Higher spin-flip modes should be rather mixed with double-exciton states $\,$\cite{foot1} and our approach becomes invalid.
Certainly, when solving the secular equation we have to limit the order of the determinant \eqref{secular} by a finite number. The latter is not a meaningful parameter in the case; we present calculation for the five-order determinant ($n_{\rm max}=m_{\rm max}=5$) noting that even the second-order one actually gives a very close result for the lowest energy. At the same time, it is important that the trivial case, when $n_{\rm max}=m_{\rm max}=1$ (i.e. the basic set is reduced to the single state$\,$\cite{ka84,di19}), results in an essentially different  spin-flip dispersion curve. In particular, then the spin-flip energy is always positive -- the gap does not vanish at any $q$.

The spin-flip spectra within the framework of our model corresponding to different electron densities $n_s(B)$ are demonstrated in Fig. 2. [The dispersion curves are not shown for $q\!<\!0.25$ since somewhere in this region  mixing with two-exciton modes becomes fairly strong $\,$\cite{foot1} (the exchange energy separating the spin-flip and magnetoplasma states becomes zero at zeroth $q$), and the applicability of the one-exciton model obviously fails.] Note that at $n_s=2.8\times10^{11}{\rm cm}^{-2}$ and $q\approx 1.4$ the spin-flip gap vanishes. This points to Stoner instability, and the calculation for lower densities/fields becomes meaningless. Experimentally the Stoner transition to the ferromagnet phase is observed at $n_s\!\approx \! 1.8\!\times\!10^{11}{\rm cm}^{-2}$.$\,$\cite{fa15,va17,va18} The dispersion curve calculated within the approximation of the single-state basis [actually a graph of the matrix element ${\cal M}_{11}(q)$] is shown in black for $n_s=2.8\!\times\!10^{11}{\rm cm}^{-2}$. Note that any value of ${\cal M}_{nn}(q)$ is always positive, and therefore an approach that ignores mixing Landau levels forbids the transition. One can see a clear difference in shape between the color curves and the black one. Thus, the model used implies existence of Stoner instability, and qualitatively correlates with the experimental data.

\begin{figure}[h]
	\vspace{-5.mm}
	\hspace{-10.mm}
	\begin{center}
		\includegraphics*[angle=0,width=.47\textwidth]{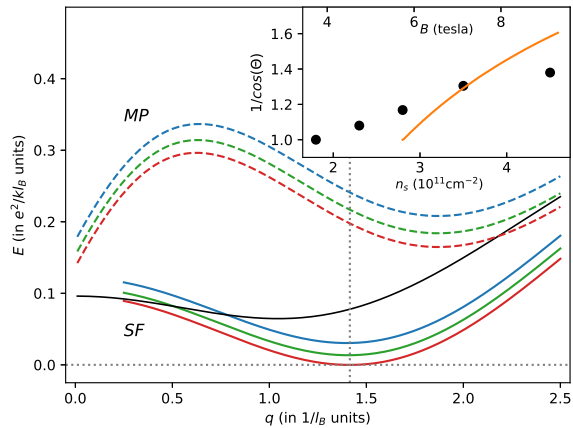}
	\end{center}
	\vspace{-7.mm}
	\caption{Spectra of spin flip (solid lines) and magnetoplasma (dash lines) modes at densities $2.8\!\times\! 10^{11}$cm${}^{-2}$ (in red), $3.6\!\times\! 10^{11}$cm${}^{-2}$ (in green) and $4.5\!\times\! 10^{11}$cm${}^{-2}$ (in blue). Black line shows the result of calculation of ${\cal M}_{11}(q)$. Inset: the inclination angle $\theta=\arctan{(B_\parallel/B)}$ at which the Stoner transition occurs, as function of density $n_s(B)$. Black circles represent experimental data.\cite{va17}}\vspace{-0.mm}
\end{figure}
For comparison we demonstrate the results for the spectra of the spinless magnetoplasma mode obtained also within the one-exciton approximation. The calculation is performed in the same way. The only difference is that now we use a basic set consisting of spin-symmetric states ${\cal R}_{{n}\,{\bf q}}^\dag|0\rangle$ where ${\cal R}_{{n}\,{\bf q}}^\dag\!\!=\!2^{-1/2}\!({\cal Q}_{0{n}\,{\bf q}}^\dag\!\!+{\cal Q}_{{\overline 0}\,{\overline n}\,{\bf q}}^\dag)$  with $n=1,2,3,...$.\cite{foot2}  We find the lowest-energy root using the equation similar to Eq. ${}\!$\eqref{secular} where now the matrix elements are ${\cal M}_{nm}^{\rm mp}=
\langle {\cal R}_{{m}\,{\bf q}}[{\cal H},{\cal R}_{{n}\,{\bf q}}^\dag] \rangle\equiv {\cal M}_{nm}$ if $n\neq m$ [see Eqs. \eqref{non_diag} and \eqref{non_diag2}], or in the $n=m$ case: \vspace{-1.mm}
 \begin{equation}\label{mp-elements}
\ba{l}
\!\!\!\displaystyle{{\cal M}_{nn}^{\rm mp}\!\!=\!nw_c\!+\!\sum_{{\bf q}_1}\frac{{\cal F}(q_1)}{q_1{\cal N}_\phi}\!\left[{h_{00}(q_1)}^2\!\!-\!|h_{0n}({\bf q}_1)|^2\right.}\vspace{-0.0mm}\\
\:{}\:{}\left.-e^{i{\bf q}\!\times{\bf q}_1}{h_{00}(q_1)}{h_{nn}(q_1)}\vphantom{{{\cal M}_{nn}(q)=\!\sum_{{\bf q}_1}\frac{{\cal F}(q_1)}{q_1{\cal N}_\phi}\!\left[|{h_{00}(q_1)}|^2\!\!-|h_{0n}({\bf q}_1)|^2\right.}}\right]\!\!+\displaystyle{{2{\cal F}(q)}|h_{0n}(q)|^2}\!/{q}\vspace{-1.mm}
\ea
\end{equation}
which specifically rewritten as
$$
\ba{l}  \displaystyle{{\cal M}_{nn}^{\rm mp}=nw_c+\int_0^\infty\!\!dq_1{\cal F}(q_1)e^{-q_1^2/2}}\mbox{\huge [}\displaystyle{1-\frac{1}{n!}\left(\frac{q_1^2}{2}\right)\!\!{\vphantom{\left(\frac{q_1^2}{2}\right)^n}}^n}\vspace{.0mm}\\\quad
\displaystyle{-L_n({q_1^2}/{2})J_0(qq_1)\mbox{\huge{]}}+\left(q^{2n-1}\!\!/2^{\,n-1}\right){\cal F}(q)e^{-q^2/2}},
\ea
$$
take the form convenient for numerical calculations.
These ${\cal M}_{nn}^{\rm mp}$ and ${\cal M}_{nm}^{\rm mp}$ expressions have to be substituted in Eq. \eqref{secular} instead of ${\cal M}_{nn}$ and ${\cal M}_{nm}$. The result is also demonstrated in Fig. 2. Note that the last term in Eq. \eqref{mp-elements} (vanishing if $q\to 0$) is precisely the term responsible for the exchange gap between the spin-flip and magnetoplasma spectra. Besides, there is $q$-independent shift $nw_c\!-\Delta_n\!=\!e_{\rm Z}$. At the same time, in accordance with the Kohn theorem,\cite{ko61} there is no Coulomb contribution to magnetoplasma energy at $q=0$.

The Stoner transition from the spin-unpolarized to the ferromagnetic state may also be provoked by an artificial increase in the Zeeman energy. This occurs when the magnetic field $B_\parallel$ parallel to the $(x,y)$ plane is applied to the system in addition to the fixed perpendicular field $B$. By ignoring any change of formfactor ${\cal F}(q)$ related to the appearance of the  $B_\parallel$  component,\cite{foot3} the recalculation of the spin-flip and magnetoplasma energies reduces simply to the $e_{\rm Z}\to e_{\rm Z}/\cos{\theta}$ replacement where $\theta$ is the inclination angle of the total field $\sqrt{B^2+B_\parallel^2}$ relative to the ${\hat z}$ direction. Within the relevant range of electron densities $n_s\!>\!2.8\!\times\!10^{11}$cm${}^{-2}$, one can find angle $\theta$ as a function of $n_s$ (or of $B$) corresponding to the vanishing spin-flip gap. See the inset in Fig. 2.


\section{Ferromagnet ground state. Discussion.}
\vspace{-2mm}

In accordance with the ideas that are similar to those used in the above approach , we model a {\em ferromagnetic} ground state considering completely occupied $(0,\uparrow)$ and $(1,\uparrow)$ sublevels, and study spin excitations from this state with the help of one-exciton basis states ${\cal Q}_{0{\overline n}{\bf q}}^\dag|0\rangle$ and ${\cal Q}_{1{\overline n}{\bf q}}^\dag|0\rangle$ ($n=0,1,2,..$). The energy of single spin-flip excitations corresponding to the $\delta S=\delta S_z=-1$ change of the spin numbers, is found from the secular equation where the diagonal elements are\vspace{-1.mm}
$$\ba{l}\vspace{-1.mm}
\!{}\!{\cal M}_{nn}^{(0)}\!\!=\!\langle {\cal Q}_{0{\overline n}}[{\cal H},{\cal Q}_{0{\overline n}}^\dag]\rangle\!=\!\Delta_n^{\rm ferr}\!+\!\displaystyle{\!\sum_{{\bf q}_1}\frac{{\cal F}(q_1)}{q_1{\cal N}_\phi}\mbox{$ \left[{\vphantom{\!\sum_{{\bf q}_1}\frac{{\cal F}(q_1)}{q_1{\cal N}_\phi}}}\right.$}|{h_{00}(q_1)}|^2}\vspace{1.mm}\\\vspace{-0.mm}{\qquad\qquad\quad\left.+\,|h_{01}({\bf q}_1)|^2
-\,e^{i{\bf q}\!\times{\bf q}_1}\,{h_{00}(q_1)}{h_{nn}(q_1)}\vphantom{{{\cal M}_{nn}(q)=\!\sum_{{\bf q}_1}\frac{{\cal F}(q_1)}{q_1{\cal N}_\phi}\!\left[{h_{00}(q_1)}^2\!\!-|h_{0n}({\bf q}_1)|^2\right.}}\right]}\!
\ea\vspace{-1.mm}
$$
($\Delta_n^{\rm ferr}=e_{\rm Z}+nw_c$) and\vspace{-1.mm}
$$\ba{l}\vspace{-1.mm}
\!{}\!{\cal M}_{nn}^{(1)}\!\!=\!\langle {\cal Q}_{1{\overline n}}[{\cal H},{\cal Q}_{1{\overline n}}^\dag]\rangle\!=\!\Delta_{n-1}^{\rm ferr}\!+\!\displaystyle{\!\sum_{{\bf q}_1}\frac{{\cal F}(q_1)}{q_1{\cal N}_\phi}\mbox{$ \left[{\vphantom{\!\sum_{{\bf q}_1}\frac{{\cal F}(q_1)}{q_1{\cal N}_\phi}}}\right.$}|{h_{11}(q_1)}|^2}\vspace{1.mm}\\\vspace{-0.mm}{\qquad\qquad\quad\left.+\,|h_{01}({\bf q}_1)|^2
-\,e^{i{\bf q}\!\times{\bf q}_1}\,{h_{11}(q_1)}{h_{nn}(q_1)}\vphantom{{{\cal M}_{nn}(q)=\!\sum_{{\bf q}_1}\frac{{\cal F}(q_1)}{q_1{\cal N}_\phi}\!\left[{h_{00}(q_1)}^2\!\!-|h_{0n}({\bf q}_1)|^2\right.}}\right]},\vspace{-1.mm}
\ea
$$
and the non-diagonal elements with ${m\not=n}$ are\vspace{-1.mm}
$$\ba{l}\vspace{1mm}
{\cal M}_{nm}^{(0)}
=\langle {\cal Q}_{0{\overline m}}{\cal H}{\cal Q}_{0{\overline n}}^\dag\rangle\vspace{1.mm}\\\qquad\qquad=-\displaystyle{\sum_{{\bf q}_1}\frac{{\cal F}(q_1)}{q_1{\cal N}_\phi}h_{00}({\bf q}_1)h_{nm}(-{\bf q}_1)e\!{}^{i{\bf q}_1\!\times{\bf q}}}
\ea
$$\vspace{-1.mm}
$$\ba{l}\vspace{0.mm}
\mbox{and}\quad{\cal M}_{nm}^{(1)}
=\langle {\cal Q}_{1{\overline m}}{\cal H}{\cal Q}_{1{\overline n}}^\dag\rangle\vspace{1.mm}\\\qquad\qquad=-\displaystyle{\sum_{{\bf q}_1}\frac{{\cal F}(q_1)}{q_1{\cal N}_\phi}h_{11}({\bf q}_1)h_{nm}(-{\bf q}_1)e\!{}^{i{\bf q}_1\!\times{\bf q}}}\,.
\ea\vspace{-1.mm}
$$
Besides, there are non-diagonal elements for which the $m=n$ case is valid: \vspace{-1.mm}
$$\ba{l}\vspace{1mm}
{\cal M}_{nm}^{(01)}
=\langle {\cal Q}_{0{\overline m}}{\cal H}{\cal Q}_{1{\overline n}}^\dag\rangle\vspace{1.mm}\\\qquad\qquad=-\displaystyle{\sum_{{\bf q}_1}\frac{{\cal F}(q_1)}{q_1{\cal N}_\phi}h_{01}({\bf q}_1)h_{nm}(-{\bf q}_1)e\!{}^{i{\bf q}_1\!\times{\bf q}}}.
\ea\vspace{-1.mm}
$$
Naturally, ${\cal M}_{nm}^{(10)}
=\langle {\cal Q}_{1{\overline m}}{\cal H}{\cal Q}_{0{\overline n}}^\dag\rangle\equiv {{\cal M}_{mn}^{(01)}}^*$.
For the case of the $6\times 6$ determinant of the secular equation (for $n,m$ running over $0,1,2$) two lowest energy-dispersion curves are shown in Fig. 3. It is convenient to identify these modes considering the $q\!\!\to\!0$ limit. In our approach the softest one (see the red line) is a spin wave ${}\!\sum_n\!\! {\cal Q}_{n\overline{n}\,{\bf 0}}^\dag|0\rangle$, i.e. $\left({\cal Q}_{0\overline{0}\,{\bf 0}}^\dag\!+{\cal Q}_{1\overline{1}\,{\bf 0}}^\dag\right)|0\rangle$. In the long-wave limit its energy
is equal to the Zeeman gap.
The blue curve corresponds to the spin-flip mode presented in the $q=0$ case by the ${\cal Q}{}^\dag\!\!\!{}_{1\overline{0}\,{\bf q}}|0\rangle|_{q\to 0}$ state and energetically shifted by $-w_c\!+\!\frac{1}{4}\int_0^\infty\!{\cal F}(p)e\!{}^{-p^2\!/2}\!{p^4}dp$ from the Zeeman level. One has to take into account that this calculation performed within the framework of our model is quite conventional -- this mode should be significantly mixed with two-exciton states, for instance, with $\large({\cal Q}{}^\dag\!\!\!{}_{0\overline{0}\,{\bf q}'}\!\!+{\cal Q}{}^\dag\!\!\!{}_{1\overline{1}\,{\bf q}'}\large){\cal Q}{}^\dag\!\!\!{}_{12\,-{\bf q}'}|0\rangle$.

\begin{figure}[h]
	\vspace{-5.mm}
	\hspace{-10.mm}
	\begin{center}
		\includegraphics*[angle=0,width=.47\textwidth]{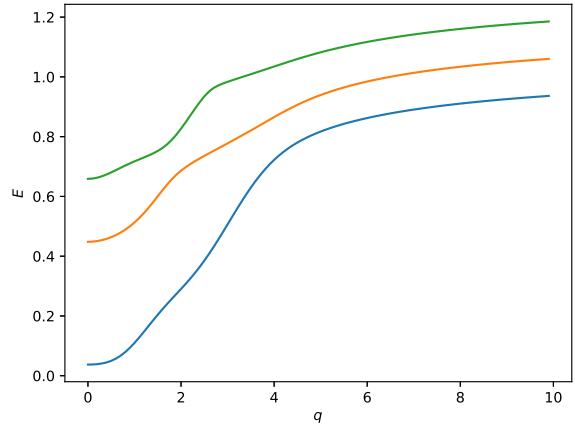}
	\end{center}
	\vspace{-7.mm}
	\caption{Spin flip spectra in the ferromagnet state (see text).}
\end{figure}\vspace{-2.mm}

The energy dispersion curves of the two lowest modes are monotonically dependent on $q$. This is shown in Fig. 3 for two electron concentrations, but the picture remains qualitatively the same throughout the range of parameters $n_s/B$ relevant for the experimental study. The corresponding gaps definitely show no tendency to vanish at any $q$. So, the studied single spin-flip excitations are obviously irrelevant to the Stoner transition. However, the $\nu\!\!=\!\!1$ QH ferromagnet is known to be very sensitive to formation of massive spin flip, for instance, skyrmion--anti-skyrmion paires for which the gap is significantly reduced with increasing parameter $r_s$ and becomes experimentally much lower than the characteristic Coulomb energy.\cite{di02} It would be natural to assume that the `reverse' Stoner transition from the $\nu\!=\!2$ ferromagnetic phase to the paramagnetic one is associated with long-wave spatial fluctuations of the spin and charge densities.  Due to the large value of $r_s$,
formation of such massive spin-flip fluctuations, presumably destroying the ferromagnet state, must occur with participation of several Landau levels. The study of this transition was not the purpose of this work~but~it~could~be~the~subject~of~future~research.

So,~using~the~excitonic~representation,~we~have considered a one-exciton model for spin-flip excitations in the $\nu\!=\!2$ QH system that is able to describe Stoner transition from the paramagnetic to the ferromagnetic phase. Our results are in a qualitative agreement with the experimental data.\cite{va17} The quantitative discrepancy is not crucial and may be due not so much to an unsuitability of the model used but to the primitive estimation of formfactor ${\cal F}(q)$.

The authors are grateful to A.B. Van'kov, I.V. Kukushkin and L.V. Kulik for useful discussions, the Russian Foundation for Basic Research for support in
performing the calculations (Grant No. 18-02-01064), and the Russian Science Foundation
(Grant No. 19-42-04119) for support in development of the Raman scattering technique and interpretation of the obtained experimental results.



\begin{thebibliography}{99}
\bibitem{foot2}\vspace{-0.mm}Spin-antisymmetric states $(\!{\cal Q}_{0{n}\,{\bf q}}^\dag\!-{\cal Q}_{{\overline 0}\,{\overline n}\,{\bf q}}^\dag)\!|0\rangle$ \vspace{-1.mm}not mixed with the ${\cal R}_{{n}\,{\bf q}}^\dag|0\rangle$ states would constitute a basis set for the $S_z=0$ component of the $S=1$ triplet.\cite{ka84,di05}
\end{thebibliography}
\end{document}